\documentclass[conference]{IEEEtran}
\IEEEoverridecommandlockouts
\usepackage{cite}
\usepackage{amsmath,amssymb,amsfonts}
\usepackage[ruled]{algorithm}
\usepackage[noend]{algorithmic}
\usepackage{graphicx}
\usepackage{textcomp}
\usepackage{xcolor}
\usepackage{forest}
\usepackage{booktabs}
\usepackage{url}
\usepackage{stfloats}
\def\BibTeX{{\rm B\kern-.05em{\sc i\kern-.025em b}\kern-.08em
    T\kern-.1667em\lower.7ex\hbox{E}\kern-.125emX}}
    
\begin{document}

\title{PExReport: Automatic Creation of Pruned Executable Cross-Project Failure Reports\\
\thanks{This work is supported by NSF Grants CCF-1846467, CCF-2007718, and CSPECC-1736209.}
}

\author{
\IEEEauthorblockN{Sunzhou Huang}
\IEEEauthorblockA{\textit{Department of Computer Science} \\
\textit{University of Texas at San Antonio}\\
San Antonio, USA \\
sunzhou.huang@utsa.edu}
\and
\IEEEauthorblockN{Xiaoyin Wang}
\IEEEauthorblockA{\textit{Department of Computer Science} \\
\textit{University of Texas at San Antonio}\\
San Antonio, USA \\
xiaoyin.wang@utsa.edu}
}

\maketitle

\begin{abstract}
Modern software development extensively depends on existing libraries written by other developer teams from the same or a different organization. When a developer executes the software, the execution trace may go across the boundaries of multiple software products and create \textit{cross-project failures (CPFs)}. Existing studies show that a stand-alone executable failure report may enable the most effective communication, but creating such a report is often challenging due to the complicated files and dependencies interactions in the software ecosystems. In this paper, to solve the CPF report trilemma, we developed PExReport, which automatically creates stand-alone executable CPF reports. PExReport leverages build tools to prune source code and dependencies, and further analyzes the build process to create a pruned build environment for reproducing the CPF. We performed an evaluation on 74 software project issues with 198 CPFs, and the evaluation results show that PExReport can create executable CPF reports for 184 out of 198 test failures in our dataset, with an average reduction of 72.97\% on source classes and the classes in internal JARs.
\end{abstract}


\section{Introduction}
\label{sec:intro}

With the assistance of third-party infrastructures and functional components, software ecosystems have expanded tremendously over the past several decades. The prevalent usage of third-party software libraries has significantly reduced the cost of software development and improved the quality of software. As the size and complexity of software products increase, developers introduce more and more software dependencies via software libraries. However, the intricate dependencies among different software projects raise new maintenance challenges. When a developer executes the software, the execution trace often goes across the boundaries of multiple software products. So, for some execution failures, it may not be easy to determine which software project should take responsibility for and fix its code. In this paper, we refer to such failures as \textit{cross-project failures (CPFs)}. For example, when a developer upgrades a software library and encounters a software failure, it may be caused by either a backward incompatibility bug in the software library or a non-robust usage of the library API.

Resolving CPFs usually requires the cooperation and negotiation of more than one project as opposed to intra-project failures (IPFs), so the failure needs to be reported from its observing software project to other projects. However, creating a good failure report is not trivial. A previous study on desirable bug reports ~\cite{bettenburg2008makes} shows that a stand-alone executable test case is the most desirable information to be included in reports. A textual failure report from client developers may not be very helpful for the library developers to investigate or reproduce the failure. Since CPFs are typically triggered at the interface between the client code and the third-party software libraries, a thorough explanation of such issues in a failure report at least needs to involve some client-side software artifacts. However, in some situations, even if the client developers provide some code, the failure can also be hard to reproduce. For example, in the comment section of an Apache Issue (Issue 2497 in Apache TIKA project)~\cite{comment}, an Apache TIKA developer complained that ``I'm not able to reproduce it w/in Solr in the unit tests with your file.''. If the failure report contains a test case that can be compiled and executed in a stand-alone way, it would greatly simplify the diagnosis process.

An ideal failure report should satisfy three essential requirements: Executability, Readability, and Conciseness. Our pilot research reveals the existence of a trilemma when client developers try to create CPF reports using existing techniques. In particular, static and dynamic slicing techniques~\cite{weiser1984program,DynaSlice}, can both prune source code based on a given \textit{seed code}, but they do not take into account the build process or the execution dependencies, so the generated slices may not be compiled or executed as a stand-alone project. Software debloating~\cite{bruce2020jshrink} is a practical technique for pruning software releases, but it focuses on destination code and execution-time dependencies instead of source code and compilation-time dependencies, so applying it to source code will cause the pruned code to be uncompilable. Finally, packaging the entire client software and sending it together as the CPF report is typically not a realistic solution, considering the numerous redundant dependencies that may (1) significantly increase the size of the report; (2) bring in lots of noise to the debugging process; (3) unintentionally leak internal information and proprietary code from other parties.

In this paper, to achieve all three requirements, we developed the PExReport, a framework for generating executable pruned CPF reports. Given a CPF, PExReport performs a three-step analysis to trace the source code and object code (e.g., Java bytecode), which were loaded at compilation and run time. In the build process, it gradually identifies all the required source code and dependencies for compiling the tests and reproduces the CPFs. Besides source code and dependencies, PExReport further identifies the required portion of build configuration, resource files, and generated source code. Finally, PExReport extracts all identified required files from the original build environment and reconstructs a stand-alone project that can compile and reproduce the CPF.

We implemented PExReport for Maven~\cite{maven} and Java combination, and performed an evaluation on 198 CPFs in 74 issues. Our results show that PExReport can reproduce 184 out of the 198 CPFs and achieve a pruning rate of 72.97\% on source classes and the classes from internal JAR files.

To sum up, this paper makes the following contributions.

\begin{itemize}
    \item A novel framework, PExReport, to extract both code and build dependencies and create pruned executable CPF reports.
    \item Three technical enhancements of PExReport to handle build configuration, resource files, and generated source code when performing automatic creation. 
    \item An evaluation of PExReport on 198 cross-project failures from 74 software project issues, showing the effectiveness of PExReport on CPF reproduction and project pruning, as well as impacts of each enhancement. 
\end{itemize}

\section{Motivation}
\label{motivation}
In this section, we conducted a pilot study in Java software ecosystems to provide real-world insights into good CPF reports and to illustrate the trilemma faced by current techniques, which motivates the design of our techniques.

\subsection{Characteristics of Ideal CPF Reports}
We conducted our pilot study using the JIRA issue tracker from Apache~\cite{apacheJIRA}. Upgrade incompatibility failures are strongly related to CPFs; therefore, we used three keywords (upgrade, incompatible, and Java) to retrieve 147 CPF reports from 32 Apache projects and tried to reproduce them.

Two researchers with more than five years of experience in software development were involved in this study. The first researcher recorded the CPF report and tried to reproduce the same failure in our build environment with the information from the report. If failure was not reproducible, the first researcher wrote down the reason, and the second researcher validated the reason. In cases where there was a conflict between the two researchers, a discussion was raised until all issues reached an agreement. Eventually, two researchers were only able to reproduce two of the 147 real-world CPF reports. We summarized the following reasons why these 145 CPF reports are difficult to reproduce; if a CPF report has more than one failed reason, we select the most direct one; the attached number is the frequency of reasons:

\begin{itemize}
\item \textbf{Never Reproduced (18)}. The client reporter provided an inaccurate test case, so the library developer never reproduced it either.
\item \textbf{Environment Specific (43)}. The failure is environment specific, so researchers cannot reproduce it without knowing the environment settings.
\item \textbf{Misuse (16)}. The client reporter misused the library, and the developer explained the reason.
\item \textbf{No Test Code (41)}. The CPF report is pure textual, and researchers failed to create a test to reproduce the same behavior.
\item \textbf{Fixed without Record (27)}. The failure was fixed without the specified fix commit, so researchers could not find the code version to reproduce it.
\end{itemize}

We believe these CPF reports do not deliver enough information for reproduction, especially for new developers who are not familiar with the software project's historical status. A previous study on desirable bug reports~\cite{bettenburg2008makes} also suggests a mismatch between what developers consider most helpful and what users provide. Based on these findings,  we summarize the following characteristics of ideal CPF reports:

\begin{itemize}
 \item \textbf{Executability.} A ideal CPF report should be easily reproduced by developers. An executable test case is the most desirable information to be included in reports. Considering the difference in build environments of client and library software projects, required code dependencies, configuration settings, and resource files should also be integrated into the test case.
 
\item \textbf{Readability.} Keeping source code accessible is essential for developers to understand the issue. With a helpful code snippet, developers could accurately identify potential misuse to address CPFs for reporters. Due to the difficulty in locating the CPFs, a good report should retain all the related source code. Destination code is not equivalent to the source code for compiled languages. For example, Java bytecode is more readable than most other destination code, but its decompiled source code with the highest ranking decompiler retains only 78\% of its semantics, according to a recent study~\cite{harrand2020java}.

\item \textbf{Conciseness.}  Considering a failure is typically triggered by a single execution, low redundancy is preferred in reporting scenarios, which could bring many benefits. A smaller report will reduce the time for network transmission and the space needed for storage (especially when there are many reports). Removing unnecessary code can significantly reduce the noise and accelerate the diagnosis progress. Furthermore, pruning redundant dependencies may also help avoid unintentionally sharing sensitive information or proprietary software artifacts.

\end{itemize}

\subsection{Cross-Project Failure Report Trilemma}

\begin{figure}[ht]
  \centering
  \includegraphics[width=.7\linewidth]{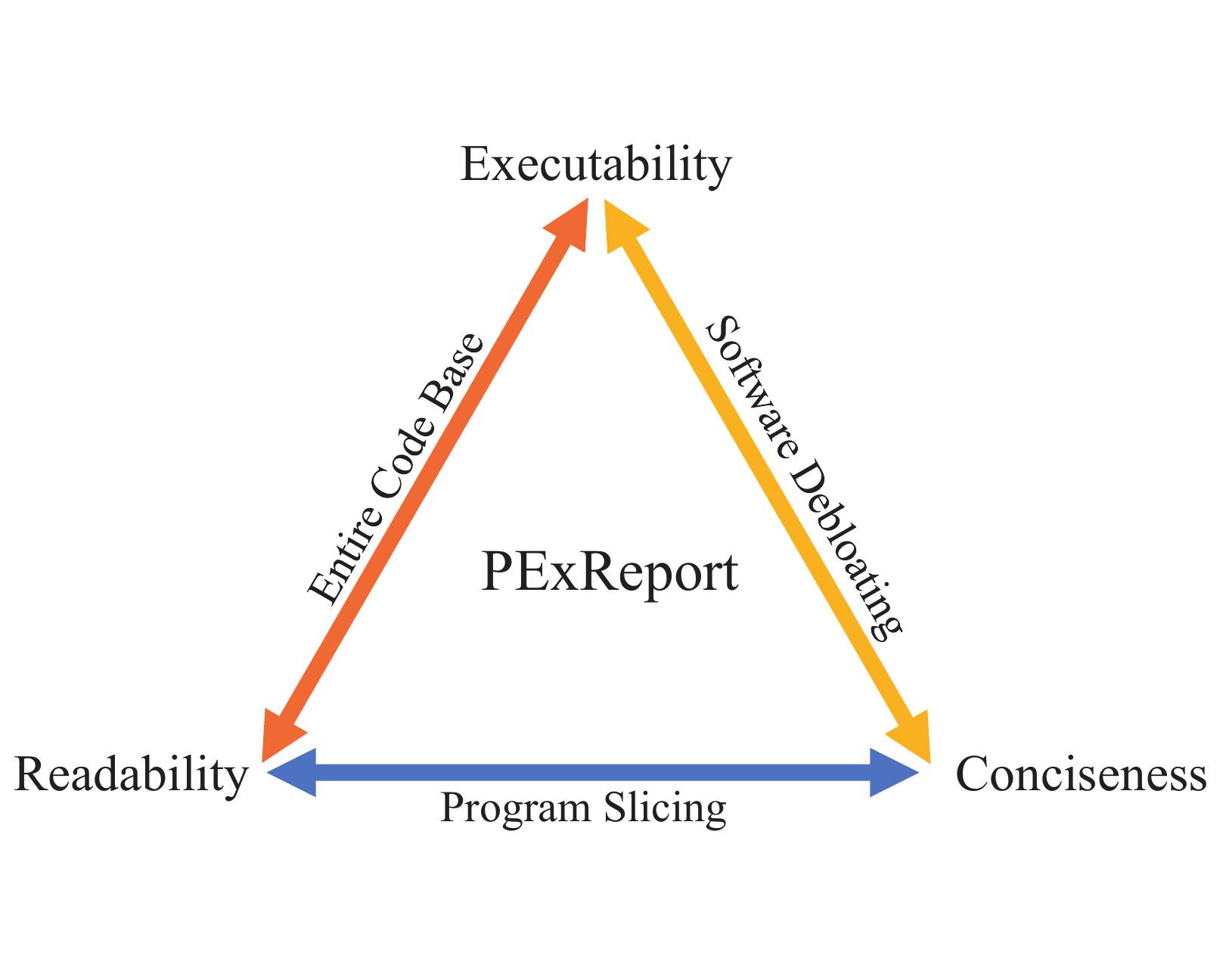}
  \vspace{-0.2cm}
  \caption{Cross-Project Failure Report Trilemma}
  \label{trilemma}
\end{figure}

To further understand the obstacles in creating CPF reports, we explored existing automatic build tools and code pruning techniques. Figure \ref{trilemma} shows the CPF report trilemma in creating ideal CPF reports with current techniques, which may partially explain why most reporters cannot create CPF reports effectively.

\textbf{Modern automatic build tools.} Maven~\cite{maven} and Gradle~\cite{Gradle} are popular tools for building and testing software in Java software ecosystems. Although both tools are designed to be platform-independent, we noticed that they lack features to extract a stand-alone test case. Using these tools can only provide an entire code base with public dependencies. Since these tools cannot identify unrelated dependencies, reporters only could attach the entire private dependencies to the test case.   In brief, modern automatic tools only achieve executability and readability but do little to reduce redundant dependencies.

\textbf{Program slicing.} Program slicing ~\cite{weiser1984program} is a dataflow-based technique for pruning source code based on a specified \textit{seed}, which is known to have limits owing to not working well with dynamic features. In addition, program slicing does not consider the execution environment or the build process. Therefore, the created slices of failures cannot be compiled or executed by themselves, which means it lacks executability.

\textbf{Software debloating.} In real-world ecosystems, debloating the size of applications is vital in embedded systems and application distribution. There are many practical tools and research studies~\cite{bruce2020jshrink} in this area. For example, ProGuard~\cite{proguard} is integrated into the Android build system, which only runs when building the application in release mode. It detects and removes unused classes, fields, methods, and attributes. In Java applications, released software does not contain source code. Because the output of compiled code is not equivalent to the source code, the developers will receive the CPF report with low readability. In short, the software debloat technique focuses on the software release stage, which does not consider the readability in the test stage.

None of the current techniques can perfectly solve this trilemma, which motivates the design of our approach to create a practical framework--PExReport, to focus on providing CPF reports with Executability, Readability, and Conciseness.

\begin{figure}[ht]
  \centering
  \includegraphics[width=.85\linewidth]{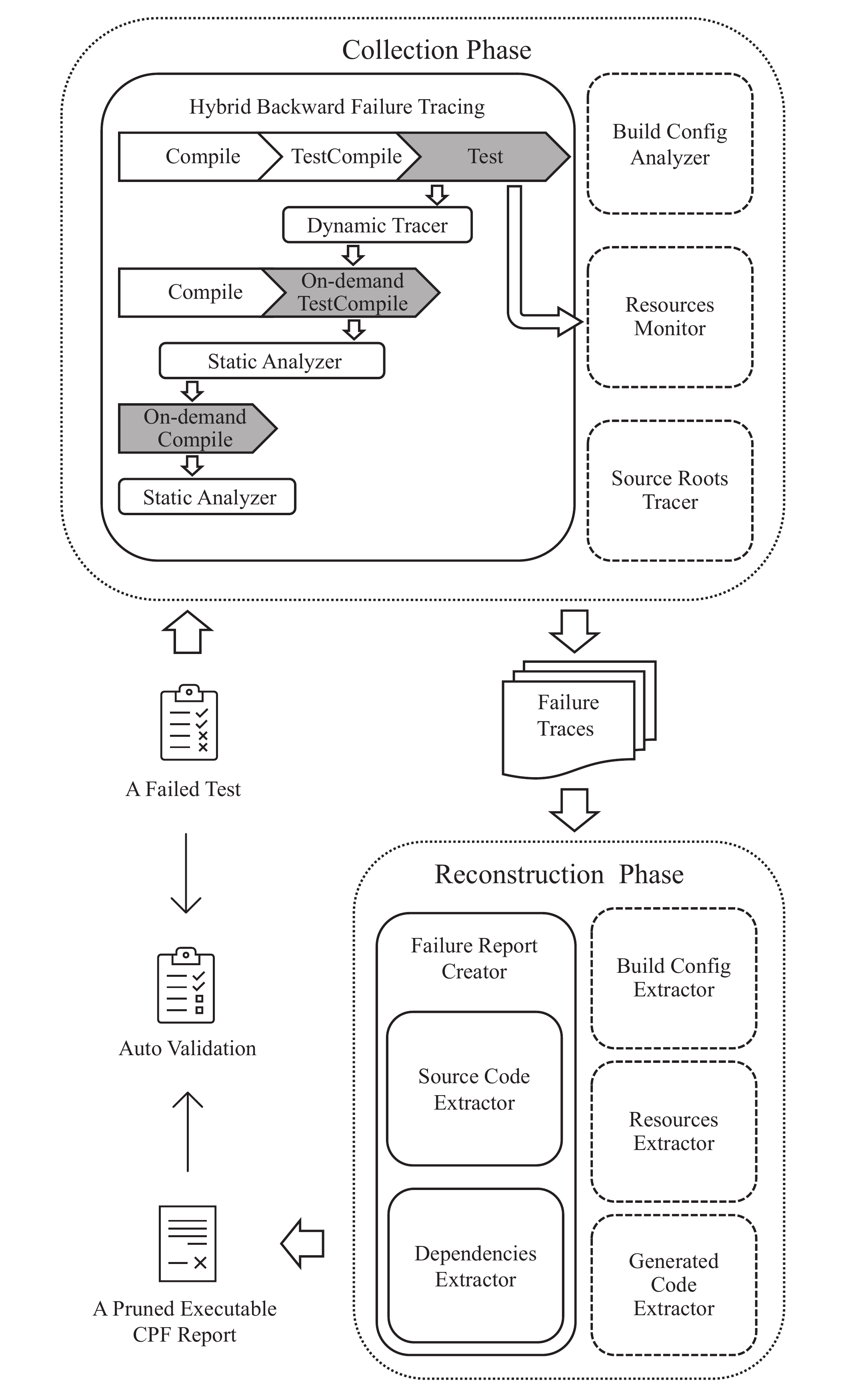}
    \vspace{-0.3cm}
  \caption{Overview of the PExReport Workflow}
  \label{overview}
    \vspace{-0.5cm}
\end{figure}
\section{PExReport}
\label{approach}

In this section, we present PExReport, a framework designed to create pruned executable cross-project failure reports automatically. Figure \ref{overview} shows an overview of PExReport’s workflow. PExReport is designed based on prevalent tasks of modern build tools involved in the lifecycle of CPFs. The input to PExReport is a failed test from an existing build environment, and the output is a pruned stand-alone executable CPF project. PExReport contains two major phases: (1) the collection phase to collect information about necessary source code, dependencies, and build environment by monitoring the underlying platforms: OS, the build tool (e.g., Maven), the compiler (e.g., Javac), and runtime (e.g., JVM), in the essential tasks of the modern build process; (2) the reconstruction phase to reconstruct a stand-alone build environment for the failed test based on the collected information. We name the collected information of a failed test from the first phase as failure traces. The arrows show the information flow between each component of PExReport. The reconstructed project for failure reporting will be automatically validated by checking whether the same error messages are triggered as in the original failure. Once the project is validated, it can serve as a reproduction package of the original failed test and will be reported as a pruned executable CPF report to the developers.

The \textit{fundamental insight} of PExReport is that it first identified and addressed the problem of pruning the build and execution environment of a test failure on top of code dependencies. This is crucial for cross-project-failure reproduction because the environment is essential for a high reproduction rate and cannot be easily shared. To overcome this problem, we develop Hybrid Backward Failure Tracing (Section~\ref{subsec:backward}) which extracts and prunes the failure by monitoring the underlying platforms in essential build tasks (i.e., \texttt{Compile}, \texttt{TestCompile}, and \texttt{Test}) of the modern build process. It takes advantage of the fact that the underlying platforms already resolve all necessary dependencies in an on-demand build process and their resolutions are the most trustworthy for reproduction. Following the same insight, we further developed three novel enhancements (Section~\ref{subsec:enhancements}) to support more heterogeneous build environment.

\subsection{Principles of Design}

PExReport should be easy to use and compatible with real-world projects and build ecosystems. We introduce the following principles to help reach the goals.  

\textbf{Source code preservation.} When clients misuse the library, a snippet of source code can be extremely helpful. Developers could also easily trace the source code to identify the fault location. Considering that even the highest ranking decompiler does not always create semantically equivalent source code~\cite{ harrand2020java}, source code preservation becomes more critical in CPF reports.

\textbf{Build environment adaptation.} A real-world project contains not only source code and dependencies; the build environment is also crucial. Modern build tools, such as Maven and Gradle, have been widely used by library developers. Given that PExReport requires CPF reports to be executable, it should be highly adaptable to modern build ecosystems. 

\textbf{Incremental creation of build environment.} As the size and complexity of projects increase, modern build tools become complex and highly customizable. Identifying the redundant build environment could be inexhaustible and hard to scale. Therefore, PExReport opts for an incremental approach that creates the build environment by identifying necessary build dependencies, as opposed to a pruning method.

\subsection{Prevalent Tasks for the Lifecycle of CPFs}
The following tasks of modern build tools form the typical lifecycle of CPFs. These tasks are generally executed in the order below, but some tasks may be omitted if not required for the current project.
\begin{itemize}
\item \textbf{Generate sources (optional).}  Generate additional source code for inclusion in compilation.
\item \textbf{Process resources (optional).}  Copy and process the resource files into the destination directory.
\item \textbf{Compile (mandatory).}  Compile the application source code of the project.
\item \textbf{TestCompile (mandatory).}  Compile the test source code, which is not coupled with the application source code.
\item \textbf{Test (mandatory).}  Run tests using a suitable testing framework; execute object code (e.g., bytecode) compiled from application and test source code.
\end{itemize}

\begin{figure}[ht]
  \centering
  \includegraphics[width=.65\linewidth]{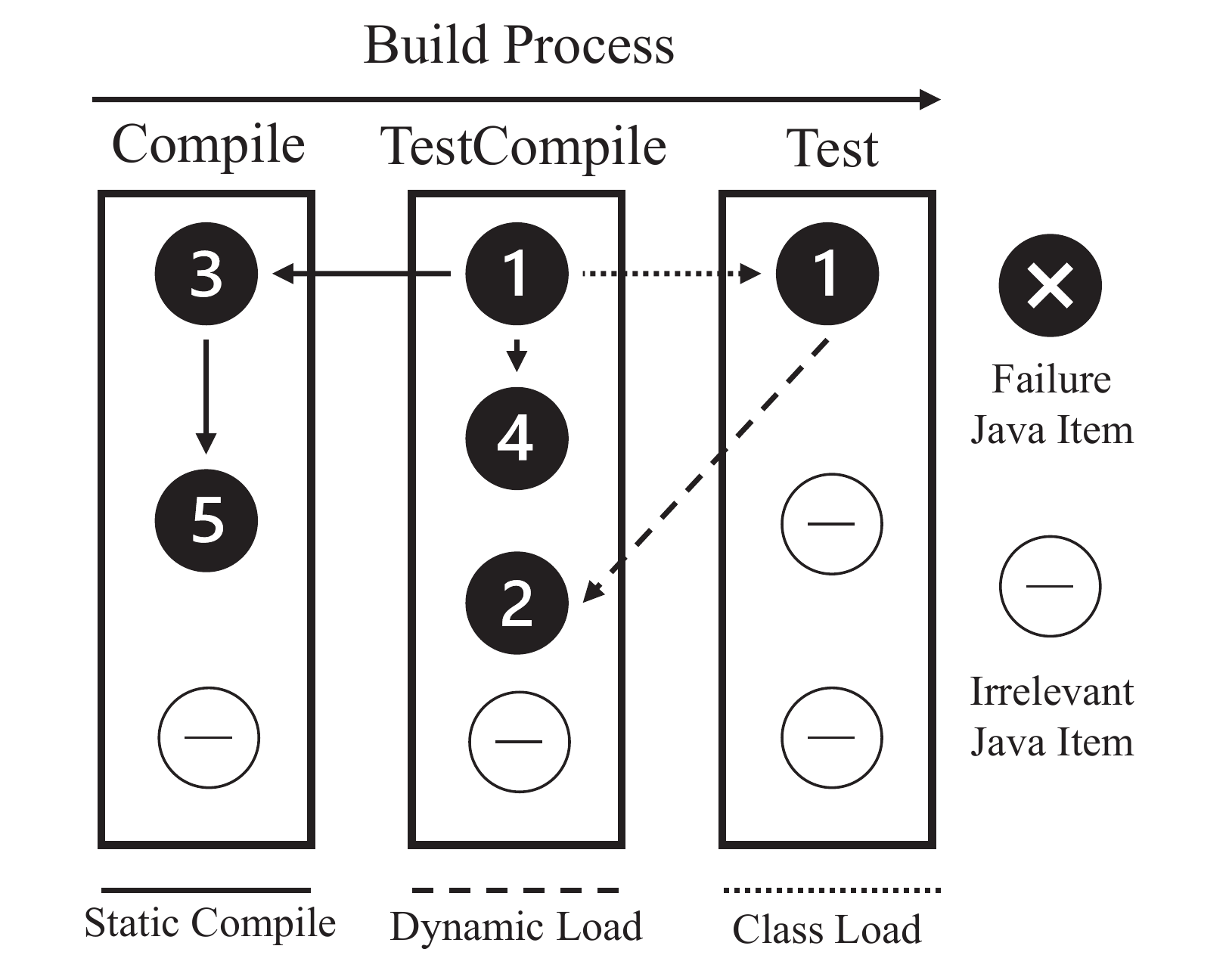}
  \vspace{-0.3cm}

  \caption{Exemplar Three Tasks Build Process of Java Project}
  \label{hbft}
  \vspace{-0.3cm}
\end{figure}

The three mandatory tasks, namely \texttt{Compile}, \texttt{TestCompile} and \texttt{Test}, are the basic steps used by build tools to reproduce a test failure. The main reason to use both \texttt{Compile} and \texttt{TestCompile} to compile source code is to make the application source code independent of the test source code. Code generation and resource management are quite popularly used in complex real-world applications. Although they are optional tasks, we should not underestimate them in reproducing real-world CPFs.

\subsection{Base Approach: Hybrid Backward Failure Tracing}
\label{subsec:backward}
In the collection phase, our base component is hybrid backward failure tracing, a three-step analysis that traces failure-related source code and library dependencies.  This component compiles and executes the failed test in its original build environment, and tracks items at \texttt{Test}, \texttt{TestCompile} and \texttt{Compile} tasks, which are essential for any failure source code to be compiled and executed.

Figure~\ref{hbft} shows the dependency tree in a build process for a failed test of Java project with three basic tasks. We refer to source code and object code (bytecode) as \textit{items}, which are used to compile and execute the failed test, respectively. All the circles are the items of the Java project, a black circle means a failure item, and a white circle means an irrelevant item. The dotted line arrow is for class loading, so the two \textit{Item 1} in the two tasks are equivalent but in a different form (source code in \texttt{TestCompile} task, bytecode in \texttt{Test} task). The solid line arrows show the static dependencies, and the dashed line arrows show the dynamic dependencies. Following the build process order, the compiled application source code in \texttt{Compile} task is provided to the subsequent \texttt{TestCompile} task as dependencies, after that, the compiled test source code of \texttt{TestCompile} is loaded to \texttt{Test} task for execution. For example, the \texttt{Compile} task compiles \textit{Item 3} (application source code), which is provided to \texttt{TestCompile} task as a reference later; the \textit{Item 1} (bytecode) of \texttt{TestCompile} task is loaded to \texttt{Test} task. The build process must be irreversible to ensure that the previous item will not depend on the later item. Given that the build process cannot predict the required items for the subsequent task, PExReport must perform the build process in three rounds to track all failure items. For example, in Figure~\ref{hbft}, the reference $1\rightarrow2$ can only be discovered in \texttt{Test} task, reference $1\rightarrow3$ can only be discovered in \texttt{TestCompile} task, and the reference $3\rightarrow5$ can only be discovered in \texttt{Compile} task. If a CPF occurs, as failures happen at the end of the build process (\texttt{Test} task), a backward tracing will be performed to obtain the failure traces. The details of Hybrid Backward Failure Tracing are described in Algorithm~\ref{alg:hbft}.

\begin{algorithm}[H]
\caption{Hybrid Backward Failure Tracing Algorithm}\label{alg:hbft}
\hspace*{\algorithmicindent} \textbf{Input:} Failed test $t$, all test source code $C_t$, all application source code $C_s$, all library object code $O_l$\\
\hspace*{\algorithmicindent} \textbf{Output:} $Traces$ 

\begin{algorithmic}[1]
 \STATE Initialize trace of $C_t$: Set $T\gets\emptyset$
 \STATE Initialize trace of $C_s$: Set $S\gets\emptyset$
 \STATE Initialize trace of $O_l$: Set $L\gets\emptyset$
 
 \STATE \textbf{Round 1:}
 \STATE \hskip1.0em  $Compile$: Object code $O_s \gets$ Compile $C_s$ by referencing $O_l$;
 \STATE \hskip1.0em  $TestCompile$: Object code $O_t \gets$ Compile $C_t$ by referencing $O_s$ and $O_l$;
 \STATE \hskip1.0em $Test(t)$: Record $R_1 \gets$ Execute $t$ by dynamic loading $O_s$, $O_t$ and $O_l$;
 \STATE \hskip1.0em $T, S, L\gets DynamicTracer(R_1);$
 
 \STATE \textbf{Round 2:}
 \STATE \hskip1.0em $Compile$: reuse $O_s$;
 \STATE \hskip1.0em $TestCompile(T)$: Record $R_2 \gets$ On-demand compile test source code in $T$ by referencing  $O_s$ and $O_l$;
 
 \STATE \hskip1.0em $T, S, L\gets StaticAnalyzer(R_2);$
 
 \STATE \textbf{Round 3:}
 \STATE \hskip1.0em $Compile(S)$: Record $R_3 \gets$ On-demand compile application source code in $S$ by referencing $O_l$;
 \STATE \hskip1.0em $S, L\gets StaticAnalyzer(R_3);$
 \RETURN $Traces\{T, S, L\}$ 
\end{algorithmic}
\end{algorithm}

In \textbf{Round 1}, PExReport runs the whole build process to the last task--\texttt{Test} with entire code base and library dependencies for both \texttt{Compile} and \texttt{TestCompile} tasks, and executes only the target failed tests in the \texttt{Test} task. PExReport uses the log of build tool to record all dynamically loaded items in the \texttt{Test} task. These items are required for the execution of the failed test in run-time, so we refer to them as $R_1$. For example, in a Java project, $R_1$ are the bytecode (object code) loaded into JVM, also referred to as Java class. The $DynamicTracer$ analyzes $R_1$, and collects the corresponding traces (usage of items).

In \textbf{Round 2}, PExReport runs the build process up to the \texttt{TestCompile} task, reusing the object code compiled from all application source code in \texttt{Compile} task. For the \texttt{TestCompile} task, PExReport on-demand compiles test source code collected from \textit{Round 1} to avoid compiling irrelevant items. In \texttt{TestCompile} task, PExReport records all referred items as $R_2$. These items are required for the compilation of test source code, so $StaticAnalyzer$ collects and combines the corresponding traces.  
    
In \textbf{Round 3}, PExReport runs a single \texttt{Compile} task of the build process, trying to compile only the application source code collected from \textit{Round 1} and \textit{Round 2}. In the \texttt{Compile} task, PExReport records all referred items as $R_3$. These items are required for the compilation of failure related application source code, so $StaticAnalyzer$ collects the corresponding traces and combines them with previous traces.

\textbf{Traces.} PExReport returns the collected failure traces of test source code, application source code and library object code in Algorithm~\ref{alg:hbft}, and sends the traces to the failure report creation component to be included in the executable CPF reports.

PExReport uses the build tools to resolve dependencies to increase robustness and non-invasiveness. Modern build tools can fetch information directly from compilers or virtual machines, meaning the resolved dependencies are the most trustworthy. The inferred dependencies from the analysis tools may be incomplete due to the quality of the tools, and some invasive tools can change the behavior of the analyzed projects. However, the build tools only provide the resolved dependencies without the dependency tree, which means that PExReport cannot use all the items in each task that may contain irrelevant items during the build process. As shown in Algorithm~\ref{alg:hbft}, to solve this problem, PExReport uses on-demand compilation\footnote{On-demand compilation, such as the implicit compilation in Java, which allows the compiler to search the required source code and dependencies to compile the designated source code.} to compile only the failure related source code and then use the references of compilation as failure traces.

The dynamic loading and static reference are handled by the run-time environment and the compiler, respectively. PExReport uses the dynamic tracer to monitor dynamic item loading in the \texttt{Test} task and the static analyzer to monitor the reference in the \texttt{Compile} and \texttt{TestCompile} tasks due to the fact that the information provided by the build tools differs between compilation and execution. The current PExReport supports the default Java compiler and multiple customized compilers, such as \textit{javac-with-errorprone}; if any compiler is not supported, PExReport only needs an adjustment to the static analyzer. As the build tools sort different types of items into different directories, the PExReport can accurately categorize items with the location information and further allow the Failure Report Creator to place them in corresponding directories.

\subsection{Three Enhancements}
\label{subsec:enhancements}

As discussed in Section~\ref{motivation}, besides source code and library dependencies, the build environment also plays an important role in the reproduction of failed tests. Although the base approach can still reproduce the failure with source code and library dependencies in the minimal standard build environment, in order to enhance the reproduction rate for real-world complex projects, we developed the following enhancements over the base components to further reconstruct a reliable build environment:

\begin{itemize}
    \item \textbf{Handling of the build configuration.} In the collection phase, the build configuration analyzer fetches the resolved build configuration and determines the necessary values of them, which will be then inserted into the template project by the build configuration extractor in the reconstruction phase.

    \item \textbf{Handling of resource files.} For the collection phase, we developed the resource monitor to watch all file accesses in the project directory during the test execution at the operating system level. Then, in the reconstruction phase, our resource extractor will copy and insert pruned resource files into the proper locations of the reconstructed project. 
    
    \item \textbf{Handling of source code generation.} Our source roots tracer tracks source code generation in the collection phase, and the generated code extractor locates the generated source code and extracts the needed source code to skip unnecessary code generation and avoid code generation conflicts.
\end{itemize}

\textbf{Handling build configuration.}
Build configuration values are necessary parts of an executable failure report. Including unnecessary configuration values will lead to more noise in the debugging phase (the debugging developer needs to consider more factors), more information leaks (e.g., internal emails and file paths), as well as potential build failures, for example, the configuration values may refer to pruned source code and dependencies which no longer exist. Therefore, we enhance PExReport to further identify and extract only build configuration values necessary for the failed test reproduction. 

In the Collection Phase, the Build Configuration Analyzer fetches the \textit{effective build configuration} (e.g., Maven provides \textit{help:effective-pom}~\cite{effective-pom}), which the build tools use to build the failed test at run time. The \textit{effective build configuration} gathers all build configuration values scattered in all build configuration files, and resolves configuration value overwriting among multiple configuration files based on the build configuration hierarchy and resolution rules, for example, Maven picks the ``nearest definition'' for resolving dependencies.
However, the \textit{effective build configuration} is still redundant for reproducing the failed test, compared with the required configuration values of the three main build phases (compile, test-compile, and test). Since tasks in the build process are performed by various plug-ins and the plug-ins can be attached to different build phases (e.g., the Java compiler plug-in can be attached to the compile and test-compile phases), PExReport further leverages the attachment relationship to identify all the plug-ins that are attached to the three basic tasks. It also excludes code-style checking and analysis plug-ins because they do not directly affect the compilation or testing. After all necessary plug-ins are identified, the Build Configuration Extractor extracts the configuration of the plug-ins from the \textit{effective build configuration} by querying with XPath. After modifying the static information, such as the absolute project path, the extractor feeds all the extracted information to the build configuration file of the reconstructed project.

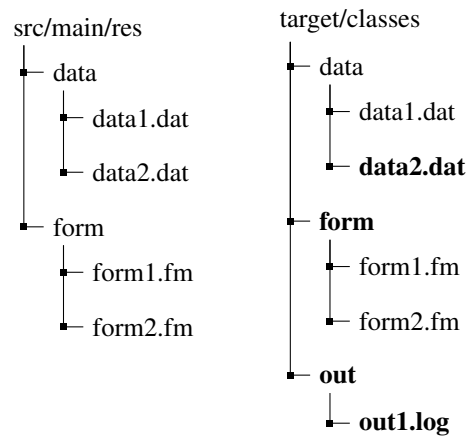
\begin{figure}[ht]
\centering
    \begin{minipage}{.4\columnwidth}
      \vspace{-1.2cm}
      \begin{forest}
  for tree={
    grow'=0,
    child anchor=west,
    parent anchor=south,
    anchor=west,
    calign=first,
    edge path={
      \noexpand\path [draw, \forestoption{edge}]
      (!u.south west) +(7.5pt,0) |- node[fill,inner sep=1.25pt] {} (.child anchor)\forestoption{edge label};
    },
    before typesetting nodes={
      if n=1
        {insert before={[,phantom]}}
        {}
    },
    fit=band,
    before computing xy={l=15pt},
  }
[src/main/res
  [data
    [data1.dat]
    [data2.dat]
  ]
  [form
    [form1.fm]
    [form2.fm]
  ]
]
\end{forest}
    \end{minipage}%
\begin{minipage}{0.4\columnwidth}
\begin{forest}
  for tree={
    grow'=0,
    child anchor=west,
    parent anchor=south,
    anchor=west,
    calign=first,
    edge path={
      \noexpand\path [draw, \forestoption{edge}]
      (!u.south west) +(7.5pt,0) |- node[fill,inner sep=1.25pt] {} (.child anchor)\forestoption{edge label};
    },
    before typesetting nodes={
      if n=1
        {insert before={[,phantom]}}
        {}
    },
    fit=band,
    before computing xy={l=15pt},
  }
[target/classes
  [data
    [data1.dat]
    [\textbf{data2.dat}]
  ]
  [\textbf{form}
    [form1.fm]
    [form2.fm]
  ]
  [\textbf{out}
    [\textbf{out1.log}]
  ]
]
\end{forest}
\end{minipage}
\caption{Exemplar Resource Directory Structure}
  \vspace{-0.4cm}

\label{fig:res}
\end{figure}

\textbf{Handling resource files.}
The Resource Monitor identifies necessary resource files for the failed test reproduction by using file system monitoring API (e.g., \textit{inotify}~\cite{inotify}--a Linux kernel subsystem that monitors filesystem events) to watch the resource file access during the \texttt{Test} phase. We use an example in Figure~\ref{fig:res} to illustrate how we monitor and extract resource files, the left side is an original resource folder existing before the build, and the right side is a target resource folder generated after the build. The files/folders accessed during testing are highlighted in bold font. 

As shown in Figure~\ref{overview}, the resource monitor only outputs files accessed at \texttt{Test} task because the \texttt{Process resources} task often copies all resource files to the target folder (e.g., the whole directory structure on the left of Figure~\ref{fig:res}). For necessary resource files accessed during the compilation process (e.g., templates of generated source code), we specially handled them by the Generated Code Extractor.
Our resource monitor also ignores all files/directories generated during the build/test process (e.g., folder \texttt{out} and file \texttt{out1.log} in Figure~\ref{fig:res}, \textit{.class} files) because these files should not be included in the reproduction package (unnecessary and causing path conflicts). For the files copied to the target location from source locations (e.g., folders \texttt{data} and \texttt{form} and all their files in Figure~\ref{fig:res}), we do not consider them as generated files, because the Resource Monitor can trace back to their source copies in the original project.

During the reconstruction phase, our Resource Extractor extracts the files and directory structure from the original project based on information collected by the Resource Monitor. For copied resource files, the Resource Extractor uses their source copies and paths tracked by the Resource Monitor (e.g., extracting \texttt{data2.dat} from the original resource folder because its copy in the target folder is accessed). Note that maintaining the structure of an accessed directory is also crucial for test reproduction. The failed test may access the directory but never access the files under it (e.g., checking the existence of a file), and some tests require a special directory structure to reproduce successfully. The Resources Extractor creates empty dummy files for the un-accessed files under the accessed directory to retain the directory structure and cover this situation. 
For example, in Figure~\ref{fig:res}, folder \texttt{form} is accessed and can be traced back to the original resource folder, but none of its files is, so the folder will be extracted, and all its files will be replaced with empty files with same names.

\textbf{Handling source roots and generated code.}
\label{subsec:generated}
The software build process may generate new source code in various ways, such as creating code from template files, generating parsing code from syntax/XML files, and even directly fetching source code from remote locations. Furthermore, code generation is often implemented in third-party tools and plugins. To handle such high flexibility of code generation in a general way, we enhance PExReport by omitting the code generation process and directly including the generated source code in the test reproduction package.

At \texttt{Generate sources} task, the build tools use source code root paths (Source Roots) to locate all (original and generated) source code. In the Collection Phase, our Source Roots Tracer tracks all the accessed source code root paths from the debug information of compilation. Next, the Generated Code Extractor utilizes the paths to identify the generated source code and excludes all original source code. In addition, the build tools may also generate some source code from code annotation processing. Our Generated Code Extractor excludes such code because it will cause compilation conflicts.

\subsection{Automatic Creation of CPF Reports}
In the reconstruction phase, our base component is failure report creation. This component creates a template project with a standard build configuration for the failure report and adds the required source code and dependencies into the project. 
 
The Failure Report Creator uses a customizable template to generate a standard project structure for the reproduction package. It uses different extractors to fetch required source code, dependencies, resource files, and build configuration. As two basic components shown in Figure~\ref{overview}, the Source Code Extractor converts the required item name to code file paths and copies these files to the generated reproduce project while maintaining the original structure; the Dependencies Extractor duplicates the local and remote dependencies and reduces the unnecessary portion based on failure traces. The generated project organizes and links the required source code and dependencies in a standard build configuration file for reproduction.
As mentioned earlier, in many cases, we also need to extract the build environment accordingly so that the replication package can successfully reproduce the failure. Therefore, the Failure Report Creator can further extract the required build configuration, resource files, and generated code by incorporating the Build Configuration Extractor, Resources Extractor, and Generated Code Extractor, respectively.

\subsection{Failure Report Validation}
After the final executable failure report (i.e., reproduction package) is constructed, the Report Validator uses a conservative strategy to ensure the report can reproduce the original test failure. In particular, a report is validated only if it generates the identical failure type and message after executing the test. Note that this strategy may reject some successfully reproduced test failures (e.g., when failure messages change with date, time, or absolute path), but it allows developers to trust the pruned failure report (and our evaluation to be conservative). Note that in practice the reporter could still manually validate the report when the automatic validation fails.

\section{Evaluation}
\label{evaluation}
This section presents our experimental results by answering the following research questions:
\begin{itemize}
\item \textbf{RQ1: How effective is PExReport in creating executable CPF reports? (Executability)}
To answer this research question, we counted the number of subjects that PExReport can exactly reproduce with the same failure type and message, and calculated the reproduction rate.

\item \textbf{RQ2: How does PExReport perform on project pruning? (Conciseness)}
To answer this research question, we calculate the reduction rates of different items from subjects and present cumulative graphs to show the performance of pruning. If a reproduction fails, we use the entire project as the CPF report (0\% reduction rate).
\item \textbf{RQ3: How effective are our techniques in PExReport on solving the CPF report trilemma?}
To answer this research question, we performed an ablation analysis using reproduction and reduction rates.
\end{itemize}

\subsection{Experiment Setup}
We implemented PExReport in Python, based on Apache Maven~\cite{maven} build tool. Since Maven is used primarily for Java, we chose Java projects as our target projects. The Archetype~\cite{archetype}, a Maven project template toolkit, is used for generating a standard Maven project for reproduction.

We performed our experiment on a Linux server running Ubuntu 16.04.5 LTS with two 8-core 2.6 GHz CPUs and 512 GB RAM. Apache Maven 3.6.3 was used to build and run tests. To avoid race conditions, we solely executed each build and test on the server. An automatic Failure Validator examined the failure types and messages to ensure the failures had been successfully reproduced. If a failed reproduction was reported, the entire project was provided as a failure report for evaluation.

\subsection{Metrics}

To answer \textbf{RQ2}, we need to use some general metrics to compare pruning performance among all subjects. Well-defined metrics could also help us to understand the results correctly. PExReport prunes the entire building environment of the failed test, including source code, dependencies, build configuration, and resource files. So, our metrics should measure the pruning on all of them and we define the following metrics:
\begin{itemize}

\item \textbf{\# Internal classes}: The number of classes from the JAR dependencies within the same organization of the subject. In the Maven convention, the same organization typically shares the same domain name in their group ID of libraries. So, we compare the group IDs to identify internal libraries and count the internal classes inside libraries.

\item\textbf{\# Source classes}: the number of classes compiled from the Java source code.

\item \textbf{\# Source+Internal classes}: The total number of source and internal classes. Our subjects have vastly different distribution strategies for main classes and internal classes. So, we combine these classes to provide a more generic metric and reduce the interference of different code distribution strategies. Since we measured the total of source and internal classes, and classes could be compressed in JARs, we chose the number of classes instead of the size of classes.

\item \textbf{Size of build configuration}: The total number of characters from the loaded POMs (Maven build configuration files), excluding the whitespace and \textit{dependencies} section for a fair comparison. We uses \textit{\# internal classes} as a metric for dependencies, which is more precise.

\item \textbf{\# Resources}: The number of resource files within the resource directories of the project.

\item \textbf{Percentage of reduction}: The percentage of reduced items, defines as the ratio of removed items to original items. For example, the percentage of reduction of internal classes is the ratio of the number of pruned internal classes to the total number of internal classes in original projects.
\end{itemize}

\subsection{Dataset Construction}
Our experiment dataset is constructed from the benchmark dataset of a research project--Sensor~\cite{wang2021will}, which triggers failures by changing dependency versions of open-source Java projects. The Sensor ground truth dataset contains 316 semantic conflicts confirmed by researchers. In our dataset, we refer to one unique failed project-library pair as one cross-project issue and one failed test case as one CPF. Although PExReport can handle CPFs raised by multiple test cases (treat multiple tests as one test), clustering tests based on dependencies is beyond the scope. Given that each issue may contain multiple CPFs, our experiment dataset consists of 74 issues with 198 CPFs. 

We do not use every CPF in Sensor dataset as subjects to evaluate PExReport because (1) some issues are irreproducible due to environmental change; (2) some CPFs are irreproducible solely since they are dependent on other CPFs (clustering non-independent tests is beyond the scope of this paper); (3) some CPFs have random factors or are flaky, so they may cause uncertainties during validation; and (4) many CPFs in one issue are identical, so using all of these repetitive failures will dominate the results. Therefore, we performed the selection for representative CPFs in the following steps:

\begin{itemize}
\item \textbf{Step 1: Verifying failed issues. }
After removing duplicated project-library pairs, we downloaded projects into our server and executed them without applying any changes. For those successfully executed projects, we re-executed them with conflicting libraries based on the Sensor dataset. Then, we selected all the issues that only failed with changed dependencies in our experiment environment. We set a 900-second time limit for each execution and removed all projects with timeout and compilation failures. Only three issues were discarded due to timeout; 119 issues were collected throughout this step.
\item \textbf{Step 2: Clustering CPFs.}
We executed all CPFs three times and removed all CPFs with inconsistent output over three executions, after which five issues were discarded.
For each of the remaining issues, we clustered CPFs with the identical failure type and message (failure group) into one cluster. As a result, we clustered 6,415 CPFs into 1,020 failure groups from 114 issues. The maximal number of failure groups within an issue is 367.
\item \textbf{Step 3: Selecting representative CPFs as subjects.}
We observed that an issue may have significantly more failure groups (i.e., 367) than others because test failures whose messages contain unique numbers (e.g., Java Object ID) are clustered into distinct failure groups. In addition, ninety percent of failure types contain no more than four failure groups.
To avoid over-representation, we further considered failure types (e.g., Assertion Failure, No Such Method) by selecting up to four failure groups at random for each failure type and issue, and then randomly selecting one representative CPF for each selected failure group. In this step, we only selected CPFs that are solely reproducible; 105 out of 395 (26.5\%) failure groups and 45 out of 238 (18.9\%) failure types have been discarded due to none of their CPFs being solely reproducible. We observed that 33 of the 107 selected issues lack internal libraries. For a more comprehensive and consistent evaluation, we only use the 74 issues with internal libraries.\footnote {The data used in this paper and PExReport implementation are available at \url{https://doi.org/10.5281/zenodo.7578677}; additional evaluation results for the 33 issues and concrete examples can be found on our website: \url{https://sites.google.com/view/PExReport/home}}
\end{itemize}

\begin{table}[htbp]
  \centering
  \caption{The Statistics of the Collected Issues.}
\begin{tabular}{lrrrrr}
\toprule
                  Item &  Min. &   Max. &   Mean &   Med. &  \textit{$\neq$0} \\
\midrule
    \# Internal classes &    32 & 36,389 &  3,848 &    955 &   74 \\
      \# Source classes &     8 &  2,457 &    265 &    144 &   74 \\
 \# Source+Internal cls &    90 & 36,417 &  4,113 &  1,331 &   74 \\
        Size of Config & 8,428 & 77,735 & 25,349 & 21,746 &   74 \\
           \# Resources &     0 &    655 &     55 &      8 &   68 \\
\midrule
       \# Total issues &       &        &           &          &       74\\
\bottomrule
\end{tabular}
  \label{tab:stat}%
\end{table}%

Eventually, we collected 198 representative CPFs from 74 issues. Table~ \ref{tab:stat} shows the statistics of the issues in our dataset. The symbol \textit{$\neq$0} means not equal to zero. The \textit{$\neq$0} column shows the count of issues that have at least one corresponding item. For instance, an issue that does not have any resource files will not be counted.. The statistics show that the issues in our final dataset have a large number of classes (4,113 Mean and 1,331 Median of \textit{\# Source+Internal classes}).

\begin{figure}[htbp]
  \centering
  \includegraphics[width=0.75\linewidth]{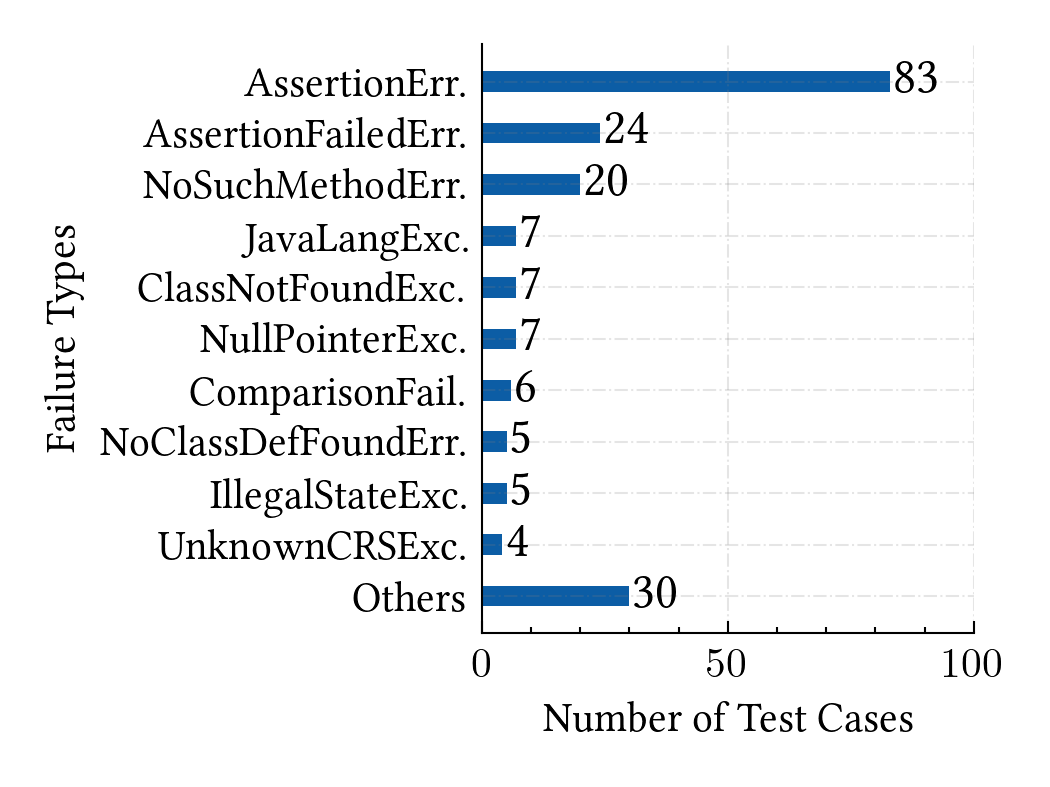}
  \caption{Top 10 Failure Types in Representative CPFs}
  \label{fig:types}
\end{figure}

We further categorized the 198 CPFs based on their failure type to show the diversity of failures in our dataset. The total number of different failure types is 31, and the top 10 failure types are shown in Figure~\ref{fig:types}. The top two failure types are assertion-related, representing more than half of CPFs. The assertion-related errors indicate that the actual variable values are not equal to the expected values. Another common failure type is ``not found''. If the dependency change modified class/method signatures, the program could still fail in the \texttt{Test} task in the case of reflection.

\subsection{Evaluation Results}

\subsubsection{The answer to RQ1}
~\\
To answer the RQ1, we executed PExReport on 198 representative CPFs and summarized the results in Table~\ref{tab:tests}. PExReport successfully reproduced 184 out of 198 CPFs with exact failure types and messages, and achieved a high reproduction rate of 92.93\%. 

\begin{table}[htbp]
  \centering
  \caption{Reproduced CPFs}
\begin{tabular}{lrrr}
\toprule
     Implementation &  \# Repro. CPFs & Repro. Rate & Perf.(sec) \\
\midrule
          \textbf{PExReport} &             \textbf{184} &      \textbf{92.93\%} &      \textbf{73.03} \\
        w/o Dynamic &              48 &      24.24\% &      66.51 \\
     Source \& Deps. &              46 &      23.23\% &      51.22 \\
   w/o Build Config &             101 &      51.01\% &      65.79 \\
   w/o Resources &             104 &      52.53\% &      68.61 \\
 w/o Generated Code &             146 &      73.74\% &      55.08 \\
\midrule
      \# Total CPFs &             198 &         &         \\
\bottomrule
\end{tabular}
  \label{tab:tests}%
\end{table}%

We investigated the 14 unreproduced CPFs to understand why PExReport cannot reproduce them. 9 of the 14 CPFs failed because of the unsupported compilers (i.e., Groovy~\cite{groovy} compiler), which could not provide the class reference information for Hybrid Backward Failure Tracing. The rest of the 5 unreproduced test cases are missing required classes at run time. We believe the Java Runtime failed to provide all the information on required classes during the test execution. The mistracking could happen when a program checks the existence of a class but never access it.

\begin{table}[htbp]
  \centering
  \caption{The Statistics of Reduction Rate}
  
\begin{tabular}{lrrrr}
\toprule
     Percent Reduction &  Min. &    Max. &   Mean &   Med. \\
\midrule
    \# Internal classes & 0.00\% & 100.00\% & 75.94\% & 84.67\% \\
      \# Source classes & 0.00\% &  99.05\% & 55.37\% & 64.39\% \\
 \# Source+Internal cls & 0.00\% &  99.65\% & 72.97\% & 79.95\% \\
        Size of Config & 0.00\% &  96.76\% & 82.96\% & 89.02\% \\
        \# Resources & 0.00\% & 100.00\% & 74.18\% & 90.91\% \\
\bottomrule

\end{tabular}
  \label{tab:red}%
\end{table}%

\subsubsection{The answer to RQ2}
~\\
We further calculated the percentage of reduction for each metric and presented Table~\ref{tab:red} to show the statistics of the percentage of reduction. PExReport performed a 55.37\% of average reduction rate on required source classes and outperformed on internal classes, source+internal classes, build configuration, and resources with 75.94\%, 72.97\%, 82.96\%, and 74.18\%, respectively.

\begin{figure}[ht]
  \centering
  \includegraphics[width=.78\linewidth]{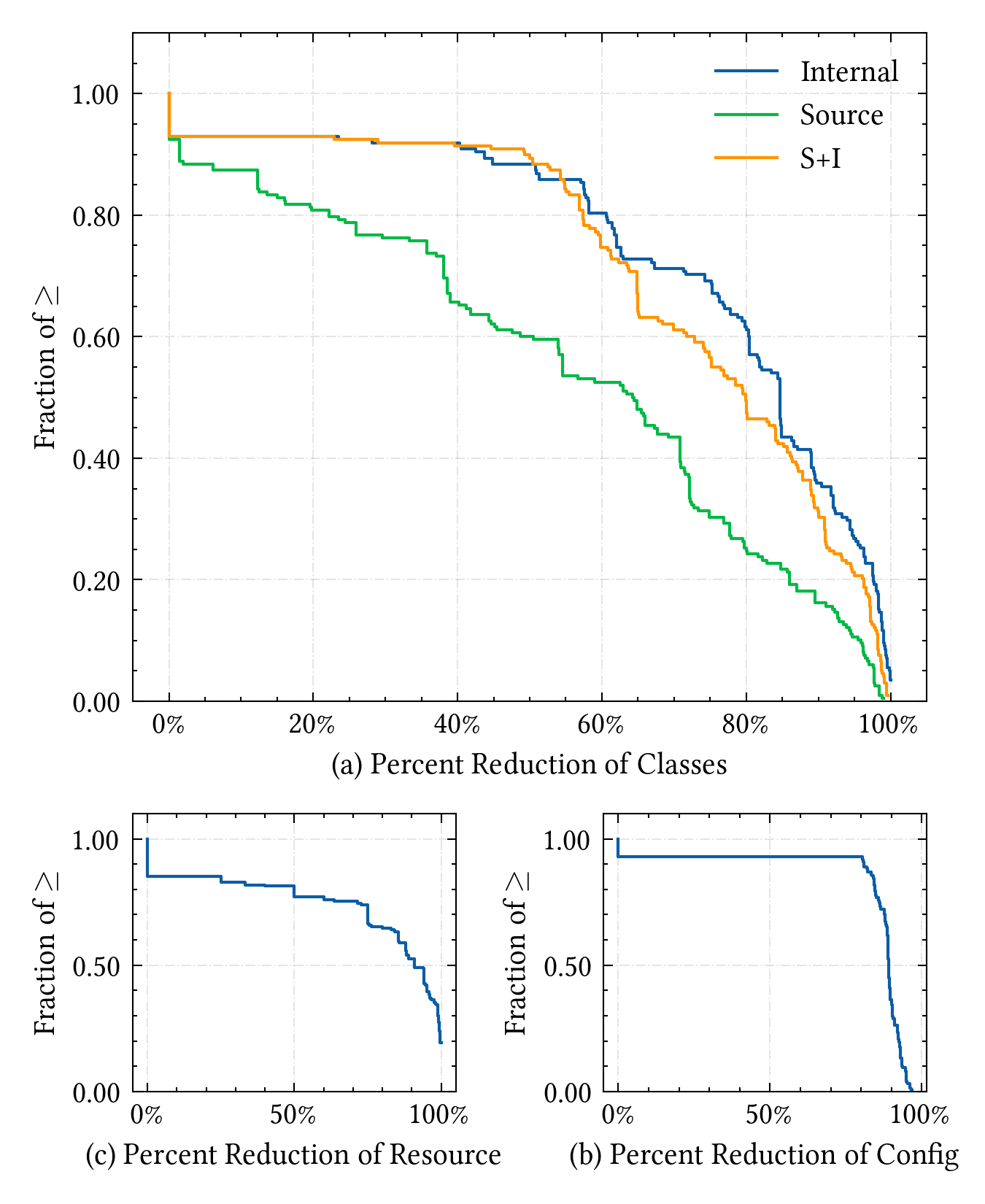}
  \caption{Cumulative Step Graph of Reduction Rates}
  \label{fig:all}
\end{figure}

We assume the reporter still wants to use the entire project as a failure report if PExReport fails to reproduce the failure. Therefore, the percentage of reduction is 0\% for unreproduced subjects. We presented the cumulative plot of results in Figure~\ref{fig:all}. The S+I represents Source+Internal, shown as the orange curve. The y-axis is the fraction of x that has satisfied the $\geq$ condition, which means that the fraction of results have at least a certain percentage of reduction. For example, a point ($x=60\%$, $y=0.79$) on the blue curve (Internal) shows that 79\% of CPFs have a reduction rate over 60\% on internal classes.

PExReport performed impressively on almost all the metrics. As the source code can provide excellent readability to developers, we believe 55.37\% of the average reduction rate is acceptable in the CPF report. Our investigation shows that the high reduction rate for build configuration is based on non-test-related configuration, such as the deployment settings and unused plugin settings. Figure~\ref{fig:all} shows that PExReport could provide a stable reduction rate for the redundant dependencies. In conclusion, the high reduction rates indicate that PExReport could provide conciseness to CPF reports.

\begin{figure*}[htbp]
  \centering
  \includegraphics[width=.9\textwidth]{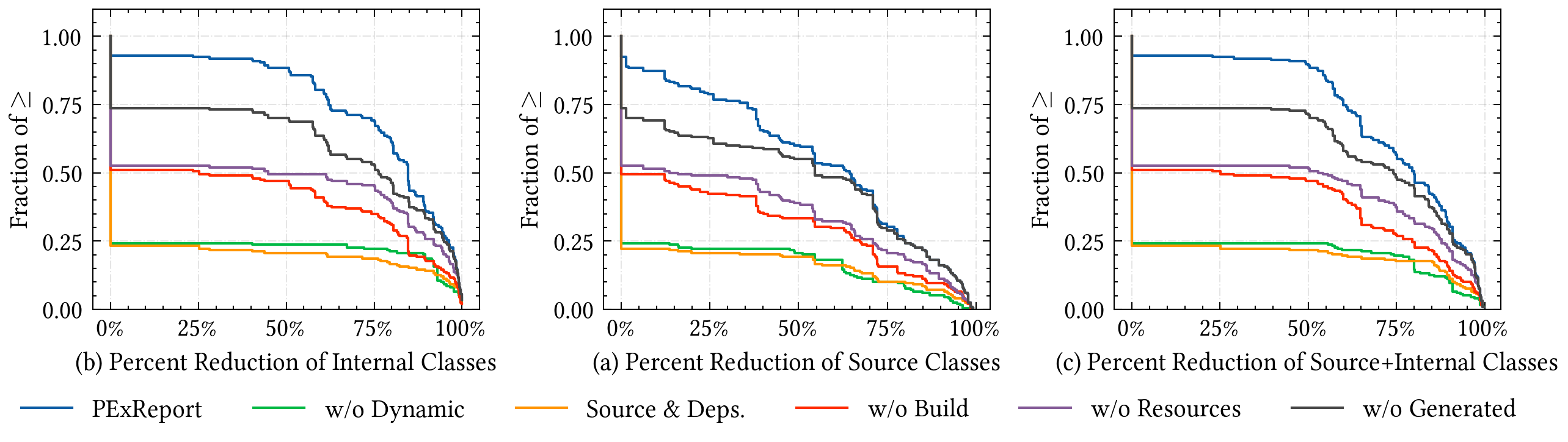}
  \caption{Comparison of Different PExReport Techniques}
  \label{fig:all-1}
\end{figure*}

\subsubsection{The answer to RQ3}
~\\
Besides the Hybrid Backward Failure Tracing, PExReport applies three enhancements to handle build configurations, resource files and generated code. We performed an ablation analysis on PExReport and calculated reproduction and reduction rates.

\textit{w/o Dynamic} represents PExReport without the dynamic tracer (JVM monitoring), which still retains the static analyzer and three enhancements. As shown in Table~\ref{tab:tests}, \textit{w/o Dynamic} only achieved a low reproduction rate of 24.24\% by reproducing 48 of 198 CPFs. Compared with the high reproduction rate of 92.93\% from PExReport, it states that pure static analysis could perform terribly in real-world Java applications as large Java applications use various dynamic features.
\textit{Source \& Deps.} represents PExReport without three enhancements, which provides source code and dependency packages with the standard Maven build configuration. The \textit{Source \& Deps.} only reached 23.23\% (46 of 198 CPFs), a low reproduction rate. Although source code and dependencies contain the most information that developers expect the reporter to provide, it is still highly likely that developers cannot use them with the standard build setting to reproduce the CPF, as exemplified in Section~\ref{motivation}. To better understand the effectiveness of the three enhancements, we turned off each enhancement one by one and compared the results with the integrated PExReport. As shown in Table~\ref{tab:tests}, the impacts of these enhancements are large. The reproduction rate reduces from 92.93\% to 51.01\% and 52.53\% without enhancements on build configuration and resource files, respectively. Without the generated code enhancement, 73.74\% of CPFs can still be reproduced because code generation is not a must for many Java projects. In contrast, developers need to deal with build configuration in every project, so the corresponding enhancement has the largest impact.

We also presented the reduction results of ablation analysis in Figure~\ref{fig:all-1}, using cumulative plots to intuitively compare our techniques.
As shown in the performance column of Table~\ref{tab:tests}, PExReport’s execution time is not heavily affected by these enhancements. As creating CPFs is not a repeated activity like the standard build process, the average reproduction finished in a minute or so should be acceptable.

Overall,  PExReport reproduced the majority (92.93\%) CPFs and provided executable cross-project failure reports with an average of 72.97\%, 82.96\%, and 74.18\% reduction rate for source and internal classes, build configuration, and resources, respectively.

\subsection{Threats to Validity}
The major internal threat to validity is the potential errors in our scripts and tool building. To reduce this threat, we carefully checked the implementation and shared them for peer review. The major external threat to validity is the variance of projects and test failures that may not be covered by our evaluation. To reduce this threat, we constructed our dataset based on the ground truth research dataset--Sensor~\cite{wang2021will} and carefully chose CPFs to avoid bias. Although PExReport is designed to work on all reproducible failures, our evaluation does not consider flaky tests and non-independent tests to avoid uncertainties and inconsistencies. Consistently reproducing and clustering failures are beyond the scope of this paper, but may also require future research.

\section{Discussion}
\label{discuss}
\textbf{Generalization to Other Languages and Building Tools.} PExReport is designed based on Maven and implemented for Java programming language. However, the idea of our approach is general and may be applied to other programming languages and build tools. The dependency analysis and enhancements for build configuration, resource files, and generated code are generalizable to most JVM languages, but compiler integration may be different from language to language, so the Hybrid Backward Failure Tracing component needs to be adjusted to support new languages. Other build tools such as Gradle / Make may allow more complicated file operations so more advanced analysis of the build process may be required.

\textbf{Duplicate Failure Reports.} PExReport takes a single CPF as its input, but one issue may generate more than one CPFs. PExReport does not resolve the potential duplication of CPFs but it can be resolved using test failure clustering and selection of representative CPF from each cluster. In our experiment, such clustering largely reduced the client CPFs to be considered. In practice, the developer may manually determine which CPF should be reported and which one should not, and select a representative CPF. Furthermore, even if the developer chose two CPFs with a similar root cause, after the pruning, representative CPFs sharing the same root cause may be reduced to similar executable CPF reports. In such cases, the client developer may need to double-check the similarity between executable CPF reports before submitting them.

\section{Related Work}
\label{related}

Though we are not aware of any research efforts creating pruned executable CPF reports to solve the CPF report trilemma, our techniques are related to existing efforts on bug reproduction, program slicing, software de-bloating, and fault localization.

\textbf{Bug and Crash Reproduction.}
Recently, JCHARMING~\cite{nayrolles2015jcharming} uses crash traces and model checking to identify program statements needed to reproduce a crash. Liu et al. proposed DoubleTake~\cite{liu2016doubletake}, which uses evidence-based analysis to largely reduce the cost of recording erroneous states. Huang and Zhang proposed LEAN~\cite{huang2012lean} to reduce the complexity of the replayed trace and the length of the replay time without losing the determinism in reproducing concurrency bugs. Weeratunge et al.~\cite{weeratunge2010analyzing} propose a novel approach that performs a lightweight analysis of a failing execution in a multi-core environment and reproduces the bug in a single-core system, under the control of a deterministic scheduler. Herbold et al.~\cite{herbold2011improved} proposed a generic and non-intrusive GUI usage monitoring mechanism to record and replay GUI bugs. Moran et al.~\cite{moran2015auto} proposed a technique that records user action steps when reproducing an Android bug, and automatically fills them into a bug report. 
Some other research efforts~\cite{hassan2018rudsea}~\cite{hassan2017automatic} try to automatically construct build or execution environments but they have not been applied to bug reproduction. Compared with existing approaches in this area, our approach focuses on the pruning and reconstruction of the build environment for buggy execution, including build configuration, resource files, and generated code.

\textbf{Program Slicing.} Program slicing~\cite{weiser1984program} is a technique to carve from a large program a smaller program that implements one or multiple features of the larger program. Program slicing often relies on code dependency graph~\cite{dietrich2008cluster}~\cite{wang2010matching} and has been used to prune a lot of targets from source code~\cite{sridharan2007thin} to pre/post conditions~\cite{harman2001overview}, paths~\cite{jhala2005path}, and databases~\cite{willmor2004program}. Dynamic slicing~\cite{DynaSlice}~\cite{DynaSlice2} identifies the statements that the buggy output depends on. Executable union slicing preserves the meaning of the original program using conditioned slicing.~\cite{executableSlices}
Compared with program slicing, PExReport focuses more on the build environment. The low reproduction rate of PExReport's variant without enhancements shows that fetching only code dependency is not sufficient. On the other hand, PExReport's enhancements can also be viewed as a slicing of the build environment, including the build configuration and the resource files.

\textbf{Software De-bloating.} Software de-bloating techniques prune a released software package to remove unnecessary dependencies and thus reduce its size, loading time, and attack surface. Pure static de-bloating techniques such as ProGuard~\cite{proguard} and Jax~\cite{tip1999practical} rely on static program slicing and more recent work such as JShrink~\cite{bruce2020jshrink} further consider runtime dependencies. Although both pruning code, compared with PExReport, the goal of de-bloating is to retain software features instead of reproducing a single execution. Also, although de-bloating may generate executable pruned code by tracking execution dependencies, it does not work on source code and does not retain the build environment.

\textbf{Fault Localization.} Statistical fault localization~\cite{tarantula}~\cite{tarantulaStudy} gives a suspicious score to each statement (or other types of code structures such as sub-control-flow-graphs~\cite{ISSTA09:BugSig}) according to the number of successful and failed test cases that cover the statement.
IR-based fault-localization~\cite{IRFaultLocal} evaluate suspiciousness scores of statements by their textual similarity to the descriptions in a given bug report.
Change-aware fault localization, such as Delta-debugging~\cite{DeltaDebug}~\cite{DeltaDebug2}~\cite{DeltaDebugReal}
considers the scenario of localizing regression faults when the history between the successful version and failed version is available. Other approaches~\cite{faulttracer}~\cite{faulttracerOld} perform impact analysis on code changes to localize faults. Hassan and Wang~\cite{hassan2018hirebuild} studied localization and repair of bugs in build scripts. On cross-project bugs, Mostafa et al.~\cite{mostafa2017experience} and Chen et al.~\cite{chen2020taming} studied behavior backward incompatibilities and their detection. Compared with all these techniques, PExReport focuses on reproduction instead of localization of bugs, so it needs to further identify and package all dependencies of a bug in its build and execution process.

\section{Future Work}
\label{future}

In the future, we plan to enhance our research in the following directions. First of all, we plan to enlarge our dataset to evaluate PExReport on more CPFs in more projects. Second, we plan to further prune the created executable CPF reports with finer-grained analysis and perform source code detailed reduction. Third, we plan to perform user studies to understand how much our tool can help developers in reproducing real-world bugs. Fourth, we plan to expand PExReport with other build tools by developing more advanced features for different software ecosystems. 

\section{Conclusions}
\label{conclude}
An executable test case is one of the most desirable features of failure reports. When reporting \textit{cross-project failures (CPFs)} to library developers, a test case is even more helpful because code is a natural way to describe interactions between library code and client code. In this paper, we developed PExReport, a framework to automatically create pruned executable CPF reports for developers, and solve the CPF report trilemma. PExReport uses Hybrid Backward Failure Tracing to identify the necessary source and dependencies, and has further enhancements to handle build configurations, resource files, and generated code. Our evaluation shows that PExReport can produce pruned executable CPF reports for 184 of 198 CPFs with an average reduction rate of 72.97\%.

\bibliographystyle{IEEEtran}
\bibliography{PExReport}

\end{document}